\documentclass[12pt]{iopart} 

\usepackage{iopams}
\usepackage{amsthm, amssymb, bm, braket}
\usepackage[dvips]{graphicx}
\usepackage{geometry}
\usepackage{xcolor, soul}
\usepackage[normalem]{ulem}

\begin{document}

\title[Role of residual interaction in the relativistic description of M1 excitation]{Role of residual \textcolor{black}{interaction} in the relativistic description of M1 excitation}

\author{Tomohiro Oishi$^{1,*}$, Goran Kru\v{z}i\'{c}$^{2,1}$, and Nils Paar$^{1,\dagger}$}
\address{$^{1}$ Department of Physics, Faculty of Science, University of Zagreb, Bijeni\v{c}ka c. 32, 10000 Zagreb, Croatia \\
$^{2}$ Research department, Ericsson - Nikola Tesla, Krapinska 45, 10000, Zagreb, Croatia}
\ead{$^*$toishi@phy.hr, $^{\dagger}$npaar@phy.hr}
\vspace{10pt}
\begin{indented}
\item[]4th June 2020
\end{indented}

\renewcommand{\bi}[1]{\ensuremath{\boldsymbol{#1}}}
\newcommand{\unit}[1]{\ensuremath{\mathrm{#1}}}
\newcommand{\oprt}[1]{\ensuremath{\hat{\mathcal{#1}}}}
\newcommand{\abs}[1]{\ensuremath{\left| #1 \right|}}
\newcommand{\crc}[1] {c^{\dagger}_{#1}}
\newcommand{\anc}[1] {c_{#1}}
\newcommand{\crb}[1] {a^{\dagger}_{#1}}
\newcommand{\anb}[1] {a_{#1}}
\newcommand{\slashed}[1] {\not\!{#1}} 

\def \beq{\begin{equation}}
\def \eeq{\end{equation}}
\def \beqa{\begin{eqnarray}}
\def \eeqa{\end{eqnarray}}

\def \bir{\bi{r}}

\def \adel{\tilde{l}} 

\begin{abstract}
{\noindent
Magnetic dipole (M1) excitation is the leading mode of multi-nucleon excitations induced by the magnetic field, \textcolor{black}{and is a phenomenon} of the spin-orbit (SO) splitting and residual interactions involved.
In this work, 
we investigate the effects of the residual interactions on the M1 excitation from a novel perspective, the framework of relativistic nuclear energy-density functional (RNEDF).
The relativistic Hartree-Bogoliubov (RHB) model is utilized to determine the nuclear ground state properties, while the relativistic quasi-particle random-phase approximation (RQRPA) is employed for the description of M1-excitation properties.
From the analysis of M1 mode in the Ca isotope chain, role of the isovector-pseudovector (IV-PV) residual interaction is discussed.
For open-shell nuclei, the pairing correlation also plays a noticeable role in the M1 mode.
The experimental data on M1 mode is expected to provide a suitable reference to improve and optimize the theoretical aspects to describe the residual interactions.
}
\end{abstract}

\maketitle

\section{Introduction} \label{Sec:intro}
Dynamics of multi-fermion interacting system represents a fundamental challenge in physics, being responsible for various phenomena in nature. 
Collective excitations in atomic nuclei represent one example, that necessitate the consideration of (i) the fermionic character of nucleons, 
(ii) effective nuclear interactions, and 
(iii) collective motion of $A$ nucleons, within a unified framework. 
One suitable method to address this interest is the quasi-particle 
random-phase approximation (QRPA) based on the energy-density functional (EDF) theory \cite{80Ring, 94BB, 03Fetter, 05BB, 1989Reinhard, 03Bender_rev, 03Dean_rev, 2005Vret}. 

\textcolor{black}{In order to understand the underlying properties of atomic nuclei, the single-particle (SP) picture within the mean-field approximation has been established. One of the fundamental properties of the SP energy levels is the spin-orbit (SO) splitting, which is essential to explain the so-called magic numbers in nuclei \cite{49Hax, 49May}.
In modern nuclear physics, one open question is how the SO splitting and the respective shell effects evolve from the valley of stability toward exotic nuclei with large neutron-to-proton number ratios \cite{2006Gaudefroy, 2011Liang, 2015Goriely}.
It is also essential in description of nuclear processes that involve unstable nuclei of relevance for nuclear astrophysics, e.g., in modeling supernova explosion and neutron-star mergers, including the r-process nucleosynthesis responsible for the production of about half of chemical elements heavier than iron \cite{2007Janka, 2011Goriely}.
}

\textcolor{black}{The magnetic dipole (M1) excitation provides one fundamental, measurable response of atomic nuclei, 
and has attracted various interests regarding the role of the SO splitting as well as the so-called residual interactions for the M1 observables \cite{2010Heyde_M1_Rev, 2008Pietralla_Rev, 2020Kruzic, 2016Birkhan, 1990Otsuka, 2005Otsuka, 1984Bohle, 1984Bohle_02, 2003Fearick, 2019Gayer, 2017Shizuma}.
In basic knowledge, the M1 transition in the leading one-body operator level would take place between the SO-partner orbits \cite{80Ring, 2010Heyde_M1_Rev, 2008Pietralla_Rev, 1978Minaev}, if the independent SP picture was a good approximation.
However, the actual M1 response is noticeably affected by the residual interactions, including e.g. the repulsive spin-isospin interaction.
This feature has been emphasized in studies based on the non-relativistic theoretical models \cite{1984Borzov, 1991Migli, 1993Kam, 2009Vesely, 2010Nest, 2010Nest_2, 2019Tselyaev}.
In Refs. \cite{2009Vesely, 2010Nest, 2010Nest_2, 2019Tselyaev} based on the Skyrme energy-density functional, for example, the difference between the particle-hole SO energies and the actual M1-excitation energies were investigated.
Note simultaneously that, however, the Skyrme-based studies have reported some ambiguities in the description of M1 properties \cite{2009Vesely, 2020Speth}. 
On the experimental side, the M1 measurement requires dedicated techniques, because of the hindrance of the M1 transitions by the other competing modes.
Nevertheless, the M1 response has been experimentally investigated using various probes, e.g., electrons, photons and hadrons \cite{1985Rich, 1990Rich, 1995Richter, 1988Lasz_M1_Exp, 1996Kneis_Exp_Rev, 2016Pai, 2019Cosel}. 
These experimental data can provide a novel reference for theoretical frameworks and effective interactions employed, which have been optimized mainly only with the ground-state properties.
}

\textcolor{black}{The purpose of this paper is to investigate the M1-excitation properties from a novel perspective, namely the relativistic nuclear energy-density functional (RNEDF) \cite{2020Kruzic, 1974Walecka, 1977Boguta}.
In comparison to non-relativistic frameworks, the relativistic one holds one advantage: 
the relativistic theory provides a natural explanation of the SO splitting in nuclei emerging from its Dirac-Lorentz formalism and degrees of freedom, which govern the interaction between nucleons \cite{1989Reinhard, 2005Vret, 1974Walecka, 1977Boguta, 2006Meng, 2016Li_32_34}. 
Considering 
the ability to naturally explain the SO splitting, its systematic application may clarify the link between the M1 response, the evolution of the SO splitting, and the residual interactions.}
Note also that, within this framework, the theory is Lorentz invariant, it obeys causality, and 
relativistic dynamics determines important phenomena in the low-energy nuclear structure. Those include the scalar and vector potentials that result in the strong SO splitting and its isospin dependence, relativistic saturation mechanisms, pseudo-spin symmetry, and nuclear magnetism in rotating nuclei \cite{1989Reinhard, 2005Vret, 2006Meng}.
The M1-based analysis may benefit these studies by optimizing the RNEDF parameters.

The paper is organized as follows.
In Sec. \ref{Sec:setting}, we present the mathematical and computational details needed for this study.
Our numerical results as well as discussions are presented in Sec. \ref{Sec:materials}.
Then in Sec. \ref{Sec:summary}, we summarize the present work.

\section{Formalism and setting} \label{Sec:setting}
We employ the CGS-Gauss system of units in this work.
Therefore, the elementary charge and nuclear magneton are given as 
$e^2 \cong \hbar c/137$ and $\mu_{\rm N}= e\hbar /(2c m_{\rm proton}) \cong 0.105$~$e\cdot$fm, respectively. 
The spherical symmetry is assumed in this study.

The RNEDF framework employed in this study is based on the relativistic four-fermion-contact interaction \cite{2008Niksic, 2014Niksic, 2020Kruzic}.
\textcolor{black}{Its effective Lagrangian density reads
\beq
  \mathcal{L} = \bar{\psi}(x) \left( i\hbar c \gamma^{\mu} \partial_{\mu} -Mc^2 \right) \psi(x) +\mathcal{L}_{\rm I}\left[\psi,~\partial \psi \right], 
\eeq
where $\partial_{\mu}=(c^{-1}\partial_{t},~{\bf \nabla})$,
$M$ is the nucleon-mass matrix, and $\psi(x)$ indicates the nucleon field.}
The exact form of the interaction term $\mathcal{L}_{\rm I}$ can be found in Refs. \cite{2014Niksic, 2008Niksic}. 
In this study, we employ the 
DD-PC1 set of the model parameters \cite{2014Niksic, 2008Niksic}. 
In a complete analogy to the meson-exchange phenomenology, 
the point coupling interaction DD-PC1 includes the isoscalar-scalar, 
isoscalar-vector, and isovector-vector channels 
with their density-dependent couplings. 
In addition, the coupling of protons to the electromagnetic field, 
and the derivative term necessary for a quantitative description of 
nuclear density distribution and radii are also taken into account. 
For the description of open-shell nuclei that necessitate 
the inclusion of the pairing correlations, 
the relativistic Hartree-Bogoliubov (RHB) model is used \cite{2005Vret, 2008Niksic, 2014Niksic, 2003Paar}.


For the particle-particle ($pp$) channel, the pairing part of the Gogny-D1S force is employed \cite{1980Gogny, 1991Berger}. 
That is, 
\beq
  V_{pp}= \sum_{i=a,b} \left\{ (W_i -H_i) +(B_i-M_i)\hat{P}_{\sigma}  \right\} e^{-d^2/\mu^2_i}, \label{eq:Vpp}
\eeq
where $d=\abs{\bir_2-\bir_1}$ is the relative distance between two nucleons, and $\hat{P}_{\sigma}$ is the spin exchange operator.
\textcolor{black}{Its parameters are given in Ref. \cite{1991Berger}.} 
This force has been utilized to reproduce the empirical pairing gaps in various nuclei. 
Note that, as defined in the original paper \cite{1991Berger}, 
the D1S force works as an attraction, but only when the 
two protons or neutrons are coupled to have $S_{12}=0$ (S0 pair). 
Thus, this pairing model is ``S0-pair promoting'', whereas 
the S1 pairing with $S_{12}=1$ should be suppressed. 
Note also that the neutron-proton pairing is neglected in this work. 

In the small amplitude limit, collective excitations can be described by the relativistic quasi-particle random-phase approximation (RQRPA). 
Since more details about this framework are given in the other publications \cite{2008Niksic, 2014Niksic, 2003Paar, 2005Niksic}, 
here we give only a brief description. 
Following the RHB solution, the quasi-particle nucleon 
operators are determined as $\crb{\rho}$ and $\anb{\sigma}$. 
Using the QRPA ansatz, 
the excited state $\ket{\omega}$ is formally given as 
$\ket{\omega}= \oprt{Z}^{\dagger}(\omega) \ket{\Phi}$, where 
$\ket{\Phi}$ is the RHB ground state of the $A$-nucleon system. 
This excitation operator reads 
\beq
  \oprt{Z}^{\dagger}(\omega) = \frac{1}{2} \sum_{\rho \sigma}
  \left\{   X_{\rho \sigma}(\omega) \oprt{O}_{\sigma \rho}^{(J,P)\dagger}
         -Y^*_{\rho \sigma}(\omega) \oprt{O}_{\sigma \rho}^{(J,P)}
  \right\},
\eeq
where $\oprt{O}_{\sigma \rho}^{(J,P)}=\left[\anb{\sigma} \otimes \anb{\rho} \right]^{(J,P)}$ is coupled to the $J^P$ spin and parity. 
Then, by solving the matrix form of the QRPA equation, 
excitation amplitudes are obtained: 
\beq
  \left( \begin{array}{cc}  A  &B  \\ B^*  &A^* \end{array} \right)
  \left( \begin{array}{c}  X^{(\omega)}  \\ Y^{(\omega)} \end{array} \right)
  =
  \hbar \omega
  \left( \begin{array}{cc}  I  &0  \\ 0  &-I \end{array} \right)
  \left( \begin{array}{c}  X^{(\omega)}  \\ Y^{(\omega)} \end{array} \right), 
\eeq
where $A$ and $B$ are the QRPA matrices \cite{80Ring, 2003Paar, 2005Niksic}. 
When the pairing correlations vanish in the ground state (GS), 
this procedure reduces to the relativistic random-phase approximation (RRPA).

\subsection{Isovector-pseudovector coupling}
For the present study of M1 excitations characterized as unnatural parity transitions, the RHB + R(Q)RPA framework needs to be further extended.
That is, in addition to the relativistic point-coupling interaction with the DD-PC1 parameterization \cite{2008Niksic}, the R(Q)RPA-residual interaction includes the IV-PV coupling term of the effective Lagrangian, 
\beq
  \mathcal{L}_{\rm IV-PV} = -\hbar c\frac{\alpha_{\rm IV-PV}}{2}
  \left[ \bar{\psi} \gamma_5 \gamma_{\mu} \vec{\tau} \psi \right]
  \left[ \bar{\psi} \gamma_5 \gamma^{\mu} \vec{\tau} \psi \right], \label{eq:AIVPV}
\eeq
where $\vec{\tau}$ indicates the isospin.
Since this IV-PV term leads to the parity-violating mean-field at the Hartree level, it does not make a contribution in the solution of natural-parity states, including the $0^+$ ground state (GS) \cite{2005Vret, 2005Niksic, 1996Podo}.
Namely, if the interest was only for the GS, this coupling would not be necessary in the RHB.
For the $1^+$ excited states by the M1 mode, however, this IV-PV Lagrangian provides a finite contribution in the R(Q)RPA residual interactions for unnatural-parity transitions: its matrix element becomes non-zero when the $1^+$ configuration is assumed \cite{2005Vret, 2005Niksic, 1996Podo}.

We mention the analogy between the present IV-PV coupling and Landau-Migdal (LM) interaction, which has been used in other theoretical studies \cite{1984Borzov, 1991Migli, 1993Kam, 2006Speth}.
The IV-PV Lagrangian relates to the spin-isospin term in the LM interaction.
More details are separately presented in \ref{App:LM_int}.

\textcolor{black}{In this work, the IV-PV coupling parameter $\alpha_{\rm IV-PV}$ is assumed to be the simple constant.
We use the value $\alpha_{\rm IV-PV}= 0.53$ fm$^2$, which is optimized by using the experimental M1 energies of $^{48}$Ca and $^{208}$Pb \cite{2020Kruzic}.
Namely, $E_{\rm exp}({\rm ^{48}Ca})=10.23$ MeV \cite{2016Birkhan, 2011Tompkins, 2017Mathy} and $E_{\rm exp}({\rm ^{208}Pb})\cong7.3$ MeV \cite{1988Lasz_M1_Exp}, respectively.
The theoretical results in terms of the centroid energy, $\bar{E}=m_1/m_0$ with $m_k \equiv \int E^k \frac{dB_{\rm M1,th}}{dE} dE$, are obtained as $\bar{E}({\rm ^{48}Ca})=9.37$ MeV and $\bar{E}({\rm ^{208}Pb})=8.02$ MeV, yielding the mean-absolute error $\cong 0.79$ MeV.
}

\textcolor{black}{One may infer an alternative form of the pseudovector coupling of the isoscalar type.
This isoscalar-pseudovector (IS-PV) coupling approximately corresponds to the spin term in the LM interaction: see \ref{App:LM_int} for details.
In Refs. \cite{1991Migli, 1993Kam} with the LM interaction, this spin term is shown to be minor compared with another spin-isospin term.
According to this analogy with the LM interaction, we neglect the IS-PV coupling in the present study.
In our results in the section \ref{Sec:materials}, the Ca isotopes are investigated, where only neutrons are active for the M1 mode.
There, we expect that dropping the IS-PV term still works as a fair approximation.
On the other side, the combined optimization of the IS-PV and IV-PV residual interactions is more complicated with multi-parameter fitting.
Further discussion on this topic is expected to be reported in future studies.
}

\begin{table}[tb] \begin{center}
\caption{Contributions by the residual interactions in 
IV-PV and pairing channels to the present RHB and R(Q)RPA calculations, 
combined with the DD-PC1 RNEDF for the particle-hole channel. } \label{table:ry832vd}
  \catcode`? = \active \def?{\phantom{0}} 
  \begingroup \renewcommand{\arraystretch}{1.2}
  \begin{tabular*}{\hsize} { @{\extracolsep{\fill}} lccc } \hline \hline
                &RHB for     &RQRPA for     &Note   \\
                &$0^+$ GS    &$1^+$ states  &        \\ \hline
  ~IV-PV        &zero        &non-zero      &particle-hole  \\
  ~Pairing      &non-zero    &non-zero, but &particle-particle, \\
  ~             &            &very small   &Gogny D1S \\  \hline \hline
  \end{tabular*}
  \endgroup
  \catcode`? = 12 
\end{center} \end{table}

In Table \ref{table:ry832vd}, the contributions of the IV-PV as 
well as the pairing interactions are summarized. 
Note that, both in the RHB and the R(Q)RPA,
the same effective interactions 
and respective parameterization, DD-PC1 (Gogny D1S), have been used  
in the particle-hole (particle-particle) channel.

\subsection{M1 response}
The M1 transitions constitute the leading mode of multi-fermion 
excitations induced by the magnetic field. 
The M1 operator reads \cite{70Eisenberg, 80Ring} 
\beq
    \oprt{P}_{\nu} ({\rm M1}) = \mu_{\rm N} \sqrt{\frac{3}{4\pi}}
    \left( g_l\hat{l}_{\nu}  +g_s\hat{s}_{\nu} \right),
\eeq
in the SP form including the spin $\hat{s}_{\nu}$ 
and orbital angular momentum $\hat{l}_{\nu}$
operators with $\nu=0$ or $\pm 1$. 
Here $\mu_{\rm N}$ is nuclear magneton, whereas 
$g$ coefficients are given as $g_l=1~(0)$ and 
$g_s=5.586~(-3.826)$ for the bare proton (neutron) \cite{70Eisenberg, 80Ring}. 
Note that the reduced matrix element of $\oprt{P}_{\nu}$ satisfies \cite{60Edm}
\beq
  \Braket{j_f(l_f) || \oprt{P}_{\nu}({\rm M1}) || j_i(l_i)} \propto  \delta_{l_f l_i}, 
\eeq
where $l$ and $j$ are the SP orbital and spin-coupled 
angular-momentum quantum numbers. 
This becomes non-zero only between the SO-partner orbits with $l_i=l_f$. 
\textcolor{black}{From this point of view, 
the M1 response seldom appears when the spin-orbit partners are both occupied or empty.}

In the present RQRPA analysis, the M1 excitations up to the one-body-operator level are considered. 
Namely, the $A$-nucleon M1 operator is given as 
$\oprt{Q}_{\nu}({\rm M1}) \equiv  \sum_{k \in A} \oprt{P}^{(k)}_{\nu}({\rm M1})$, 
where $\oprt{P}^{(k)}_{\nu=0,\pm1}$ is the SP-M1 operator of the $k$th nucleon. 
Its strength is obtained as 
\beq
  \frac{dB_{\rm M1}}{d E_{\gamma}} = \sum_{f} \delta(E_{\gamma}-\hbar \omega_f) \sum_{\nu} \abs{\Braket{\omega_f | \oprt{Q}_{\nu}({\rm M1}) |\Phi}}^2, \label{eq:dis_BM1}
\eeq
for all the positive R(Q)RPA eigenvalues, $\hbar \omega_f >0$. 
For plotting purpose, this discrete strength is smeared with the Cauchy-Lorentz profile of the full width at half maximum, $\Gamma_{\rm FWHM}=1.0$ MeV. 
Note that, in this work, we neglect the effect of the meson-exchange current as well as the couplings of configurations \cite{2010Heyde_M1_Rev, 1990Richter, 1991Migli, 1993Kam, 1977Dehesa, 1988Takayanagi, 2008Marcucci, 1994Moraghe, 1982Bertsch, 2006Ichimura}, which need further multi-body operations going beyond our present scope.

\begin{figure}[t] \begin{center}
  \includegraphics[width = 0.6\hsize]{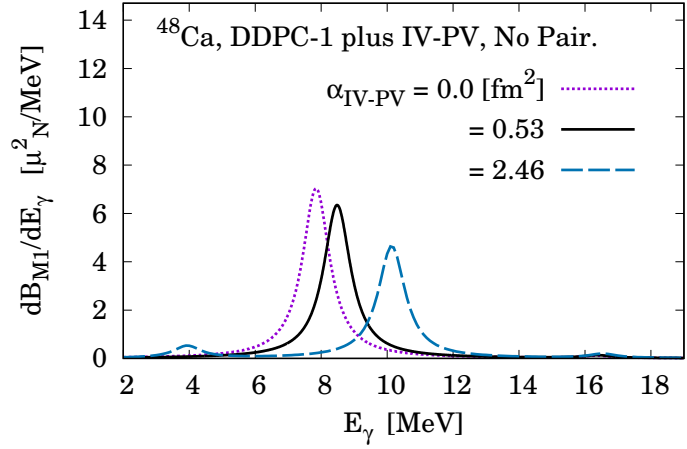}
  \caption{\textcolor{black}{M1 strength of $^{48}$Ca calculated with 
several IV-PV coupling coefficients. 
The discrete strength in Eq. (\ref{eq:dis_BM1}) is smeared with the Cauchy-Lorentz profile 
of the full width at half maximum, $\Gamma_{\rm FWHM}=1.0$ MeV. }
} \label{fig:6651}
\end{center} \end{figure}

\section{Results and discussions} \label{Sec:materials}
In the following, we show the results for the M1 transitions of the even-even $Z=20$ (Ca) isotopes from the $0^+$-ground to the $1^+$-excited states.
With the DD-PC1 and D1S parameterization used for the 
RNEDF and the pairing correlations, respectively, 
we confirmed that the particle-bound systems in the ground
state are obtained up to $N=16$-$44$ for Ca isotopes. 
Our RHB plus R(Q)RPA calculations are performed based on the harmonic-oscillator basis up to the $20$ major shells.
The cutoff energies of the RQRPA configuration space are fixed to provide a sufficient convergence of the M1-excitation strength \cite{2020Kruzic}.

\subsection{Role of IV-PV interaction} \label{Sec:IVPV_Test}
Before going to the systematic calculations, we check the effect of the IV-PV interaction on the M1 excitations.
For this purpose, we choose the $^{48}$Ca nucleus for benchmark, because the pairing correlations vanish in its GS, and thus, the IV-PV effect is purely seen.

Figure \ref{fig:6651} demonstrates the evolution of the M1 response for a variation of the IV-PV coupling coefficient, 
$\alpha_{\rm IV-PV}$, given in Eq (\ref{eq:AIVPV}). 
For comparison, the experimental M1-excitation energy 
of the $^{48}$Ca nucleus is also displayed: $E_{\gamma}=10.23$ MeV for M1 ($0^+ \rightarrow 1^+$) \cite{2016Birkhan, 2011Tompkins, 2017Mathy}. 
From this result, one can read that the larger $\alpha_{\rm IV-PV}$ value leads to the higher M1-excitation energy.

As mentioned in the previous section, the $\alpha_{\rm IV-PV}$ is adjusted to minimize the gaps between the theoretical and experimental M1-excitation energies for the $^{48}$Ca and $^{208}$Pb nuclei.
As the result, $\alpha_{\rm IV-PV}=0.53$ fm$^2$ is determined as the best fit throughout the light to heavy-mass regions: see also Ref. \cite{2020Kruzic}.
On the other hand, if we insist in reproducing the M1-excitation energy of $^{48}$Ca, \textcolor{black}{$E_{\gamma}=10.23$ MeV \cite{2016Birkhan, 2011Tompkins, 2017Mathy}}, $\alpha_{\rm IV-PV}=2.46$ fm$^2$ is necessary as displayed in Fig. \ref{fig:6651}, but is less adequate for the heavier system $^{208}$Pb.
This ambiguity possibly originates in the simple-constant coupling.
One may consider more advance but complicated parameterization for $\mathcal{L}_{\rm IV-PV}$ to improve the consistency with the measured M1-excitation data.
In this work, however, we hold the original setting as in Ref. \cite{2020Kruzic}, in order to avoid the confusion.

\begin{figure}[t] \begin{center}
  \includegraphics[width = 0.6\hsize]{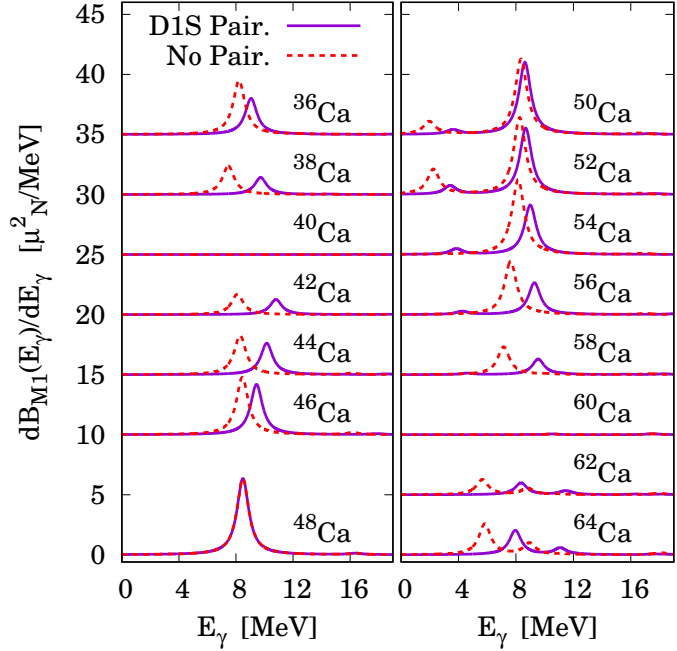}
  \caption{The M1 ($0^+\rightarrow 1^+$) transition strength distribution for Ca isotopes obtained with the DD-PC1 functional. 
Results with the Gogny-D1S pairing interaction (solid line) and without pairing correlations (dotted line) are separately shown.} \label{fig:1123}
\end{center} \end{figure}

\subsection{Isotopic evolution}
Figure \ref{fig:1123} shows the M1-transition strength distributions 
from our systematic calculations for the $^{36-64}$Ca isotope chain. 
First we focus on $^{40}$Ca, where the M1 strength almost vanishes. 
This is simply because no SO-partner orbits 
are available for the M1 transition: in the ground state of $^{40}$Ca, 
all the $(1p_{3/2}~\&~1p_{1/2})$ and $(1d_{5/2}~\&~1d_{3/2})$ 
orbits are fully occupied, and thus, 
the M1 transitions between these orbits are forbidden. 
The allowed transition is e.g. from the bound $1d_{5/2}$ 
to the high-continuum $d_{3/2}$ orbits. 
However, these transitions are strongly suppressed because 
the overlap of their radial wave functions is small. 
The same feature also appears for $^{60}$Ca with 40 neutrons, where up to the $2p_{1/2}$ and $1f_{5/2}$ single-particle states all neutron orbits are occupied.
Consequently, up to the one-body-operator analysis, the nucleon numbers 20 and 40 are the ``M1-silence'' points.
\textcolor{black}{We notify that, for $^{40}$Ca, its M1 strength was indeed experimentally found near $10$ MeV, and the theoretical calculation with some extensions beyond the standard QRPA reproduces this strength \cite{1989Kamerd}.
}

As shown in Fig. \ref{fig:1123}, the evolution of M1 response along the $^{36-64}$Ca 
isotope chain results in one remarkable peak in each system. 
This is attributable to the M1 excitation of valence neutrons, whereas 20 protons are M1-silent. 

\begin{figure}[tb] \begin{center}
  \includegraphics[width = 0.6\hsize]{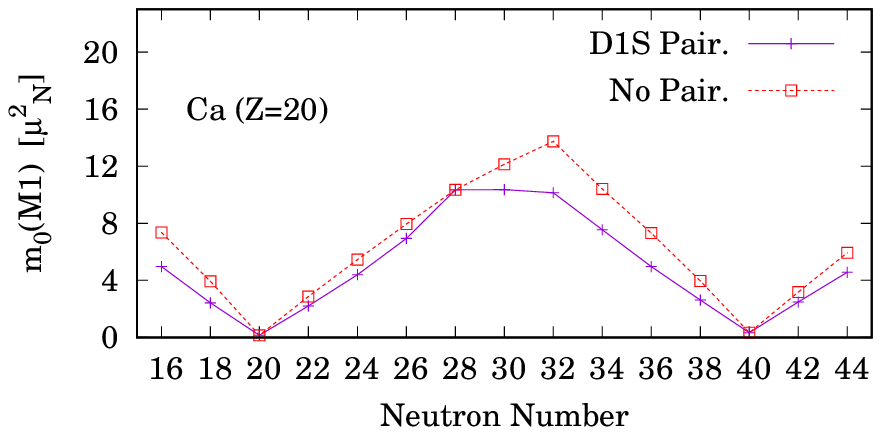}
  \includegraphics[width = 0.6\hsize]{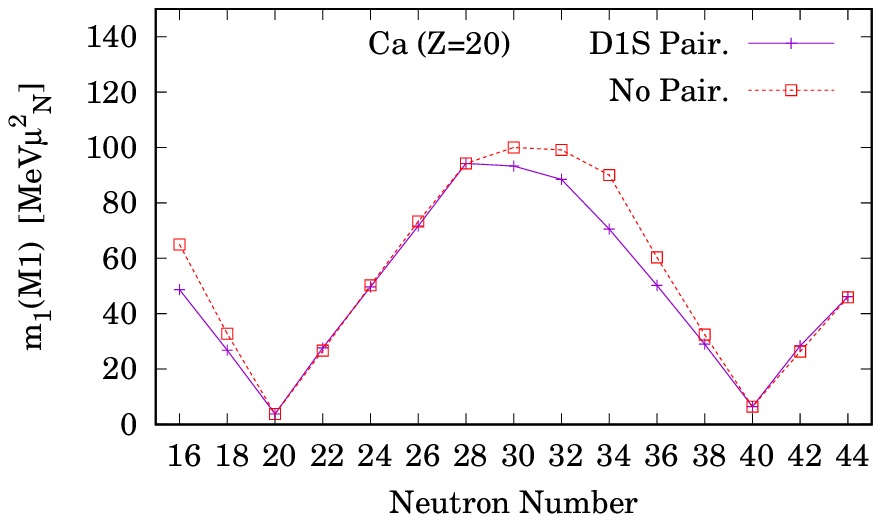}
  \caption{Summations of the M1-excitation strength of Ca isotopes, $m_k \equiv \int E^k \frac{dB_{\rm M1}}{dE} dE$.
(Top panel) Non-energy-weighted sum with $k=0$.
(Bottom panel) Energy-weighted sum with $k=1$.
} \label{fig:1019}
\end{center} \end{figure}

We also consider the case where the pairing correlations are not taken into account, as shown in Fig. \ref{fig:1123}.
In this setting, there are no mixtures of different configurations in the M1 states.
For the $^{36-38}$Ca, $^{42-58}$Ca, and $^{62-64}$Ca nuclei, the no-pairing M1 response is explained purely from the neutron transitions of
$(1d_{5/2}\rightarrow 1d_{3/2})$, 
$(1f_{7/2}\rightarrow 1f_{5/2})$, and 
$(1g_{9/2}\rightarrow 1g_{7/2})$, respectively. 
This behavior is indeed expected from the ordering of the nuclear-shell orbits.
For the $^{50-52}$Ca isotopes, the second, low-energy peak appears 
due to the $(2p_{3/2}\rightarrow 2p_{1/2})$ transition. 
Notice also that, since the higher SO-partner orbits, 
$1f_{5/2}$ and/or $2p_{1/2}$, are occupied in $^{54-60}$Ca, 
the M1 response is consistently reduced by these blocking-neutron states. 

When the D1S-pairing interaction is included in the calculations, 
the M1 transition strength becomes reduced, as shown in Fig. \ref{fig:1123}. 
This is understood from the S0-pair promoting ability of the D1S force.
When the S0-pair component is dominant in the ground state, its M1 response is suppressed \cite{2019OP}.
\textcolor{black}{See \ref{App:KYVUT} for details.}
Also, the pairing correlations invoke the mixture of different SO-partner transitions.
For example, in $^{42-46}$Ca, we confirmed that the dominant component is still $(1f_{7/2}\rightarrow 1f_{5/2})$, but simultaneously the component $(1d_{5/2}\rightarrow 1d_{3/2})$ has finite contribution in the main M1 peak. 
This is a result of the smearing of the Fermi surface in 
the RHB solution due to the pairing correlations, 
and thus, the $1d_{3/2}$ state is not fully occupied. 

\subsection{Sum-rule investigation}
Figure \ref{fig:1019} shows the sum of the M1 transition strength for Ca isotopes. 
That is, 
\beq
   m_k ({\rm M1}) \equiv \int E^k_{\gamma} \frac{dB_{\rm M1}}{dE_{\gamma}} dE_{\gamma}.
\eeq
The results show a strong dependence of the $m_0$ value on the M1-active nucleons, supported by the analysis of relevant two-quasi-particle configurations in the main M1 peaks. 
First, we focus on the case without the pairing correlations.
For $^{40-60}$Ca, the M1 excitations are dominated by the transitions of $(1f_{7/2}\rightarrow 1f_{5/2})$ and $(2p_{3/2} \rightarrow 2p_{1/2})$. 
Thus, the $m_0$ value simply increases or decreases according to the interplay of active and blocking neutrons in these orbits. 
Second, when the D1S pairing correlations are included, the $m_0$ value is commonly reduced in comparison to the no-pairing result. 
This behaviour is consistent to the strength distributions shown in Fig. \ref{fig:1123}. 
The reduction of the M1-sum value is understood from the S0 component in the ground state, which is enhanced by the D1S-pairing force \cite{2019OP}.
See also \ref{App:KYVUT}.

\textcolor{black}{It is worthwhile to mention the Kurath M1-sum rule \cite{1963Kurath}.
By adjusting it to our convention of units, as concluded by Kurath, the energy-weighted M1 summation approximately satisfies that,
\beq
  m_{k=1}({\rm M1,~Kurath}) \cong -E_{\rm SO} \frac{3}{4\pi} \left( g^{\rm IV}_s + \frac{1}{2} \right)^2 \mu^2_{\rm N} \sum_{i} \Braket{\bi{l}(i) \cdot \bi{s}(i)}, \label{eq:Kurath}
\eeq
where $g^{\rm IV}_s=-4.706$, the bracket means the expectation value for the ground state, and the summation $\sum_{i}$ is only for the M1-active nucleons.
The attractive $E_{\rm SO}<0$ indicates the general, non-relativistic spin-orbit energy as determined in the Eq. (2) in Ref. \cite{1963Kurath}.
Therefore, as long as the M1-active nucleons are in the common orbit with the same $\Braket{\bi{l}(i) \cdot \bi{s}(i)}$ value, the energy-weighted M1-sum rule is simply proportional to the number of those nucleons.
}

\textcolor{black}{In the bottom panel of Fig. \ref{fig:1019}, we plot the energy-weighted summation, $m_1({\rm M1})$, obtained with the present R(Q)RPA method.
For $^{40-48}$Ca as the best example, the $m_1({\rm M1})$ value shows a linear increase to the neutron numbers, and thus, it agrees with the Kurath sum rule.
For these nuclei, the M1 response is mostly attributed to the $(1f_{7/2}\rightarrow 1f_{5/2})$ transition independently of whether the pairing interaction is active or not.
We checked that the SO-gap energy between the $(1f_{7/2})$ and $(1f_{5/2})$ levels is roughly $8$ MeV with a small fluctuation for $^{40-48}$Ca in the RHB results.
This is equivalent to $E_{\rm SO}\cong -2$ MeV in terms of the Kurath's formalism.
Remember also that $\Braket{\bi{l}(i) \cdot \bi{s}(i)}=3/2$ commonly for the M1-active $(1f_{7/2})$ neutrons.
The prediction by Eq. (\ref{eq:Kurath}) with these quantities yields that $m_1({\rm M1,~Kurath}) \cong 12.6\cdot N_{\rm M1}$ MeV$\mu^2_{\rm N}$, where $N_{\rm M1}$ indicates the number of M1-active neutrons.
Namely, $m_1({\rm M1,~Kurath}) \cong 25.2$, $50.4$, $75.6$, and $100.8$ MeV$\mu^2_{\rm N}$ for $^{42-48}$Ca, respectively.
In comparison, our actual $m_1({\rm M1})$ values are obtained as $27.6$, $49.7$, $71.7$, and $94.2$ MeV$\mu^2_{\rm N}$ for $^{42-48}$Ca, respectively.
These values are well consistent to the prediction by Eq. (\ref{eq:Kurath}).
However, we simultaneously notify that, in the RNEDF framework, there is not a corresponding parameter to $E_{\rm SO}$ in Ref. \cite{1963Kurath}.
Instead, the SO splitting is concluded from the competition of two independent ingredients, namely, the scalar and vector potentials in the relativistic mean-field calculations \cite{1989Reinhard, 2005Vret, 1974Walecka, 1977Boguta}.
Nevertheless, up to the one-body QRPA level, the Kurath sum rule is well reproduced in certain nuclei within the RNEDF framework.
This result supports a validity of the RNEDF calculation applied to the M1 mode.
}


\textcolor{black}{In the no-pairing case, we checked that our $m_1({\rm M1})$ results in Fig. \ref{fig:1019} are consistent to the Kurath sum rule, for which $E_{\rm SO}$ is fixed to mimic the respective SO-gap energy.
Then, when the pairing interaction is switched on, our $m_1({\rm M1})$ values differ from the Kurath sum rule, e.g. for $^{50-58}$Ca.
In these systems, two SP orbits, $(1f_{7/2})$ and $(2p_{3/2})$, are both relevant in the RHB-GS solutions.
The pairing correlation invokes a mixture of these orbits, and thus, the condition for the Kurath sum rule is not complete.
}

\begin{figure}[tb] \begin{center}
  \includegraphics[width = 0.57\hsize]{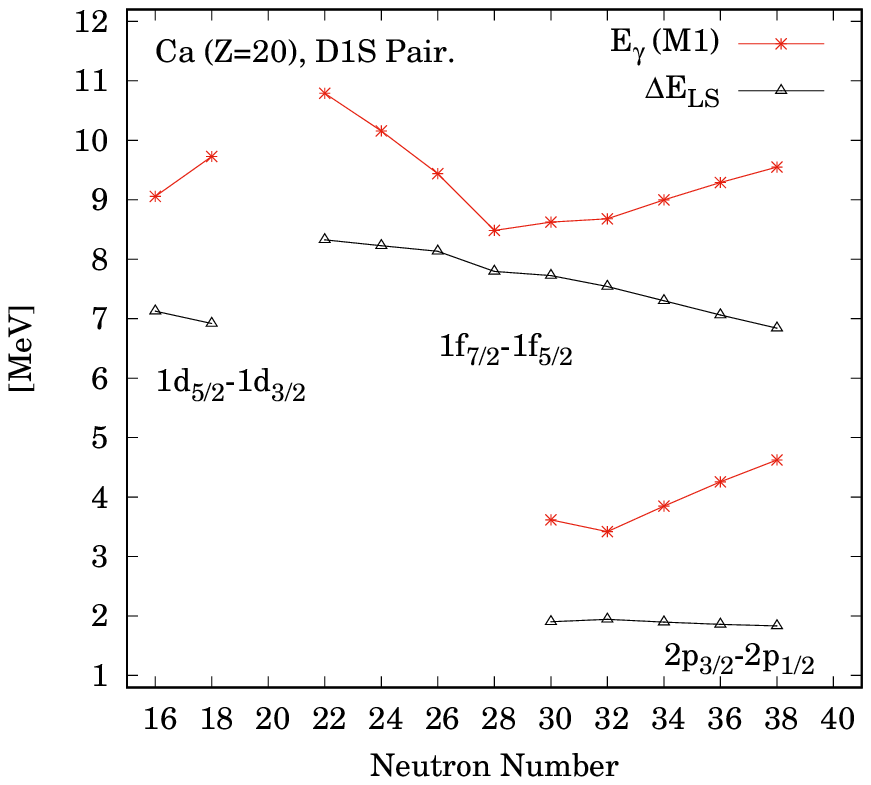}
  \includegraphics[width = 0.57\hsize]{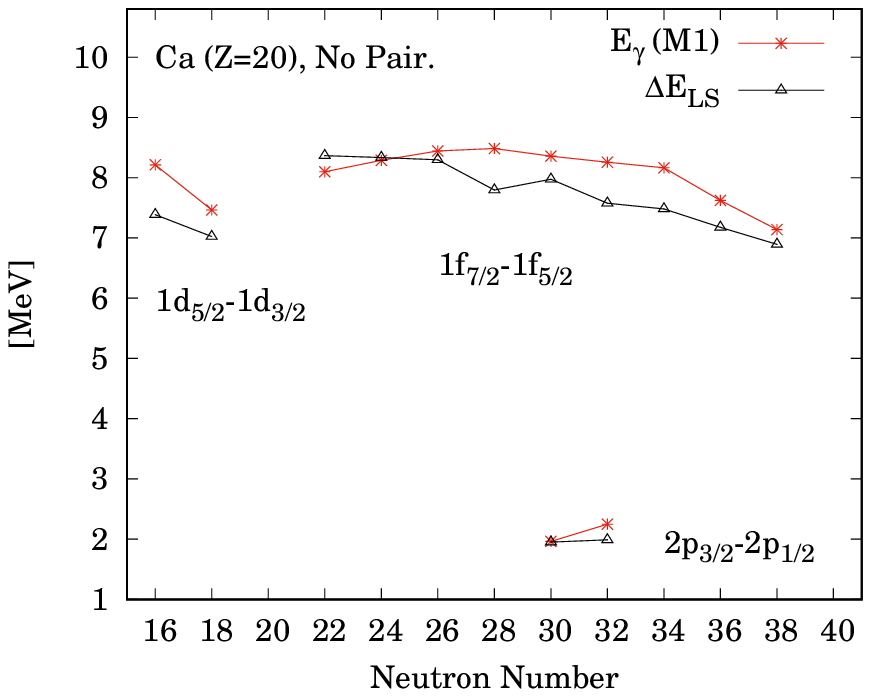}
  \caption{(Top panel) The M1-excitation energies $E_{\gamma}({\rm M1})$ of Ca isotopes solved with our R(Q)RPA procedure, and the corresponding SO-gap energies $\Delta E_{\rm LS}$ in their RHB solutions. 
The calculations are based on the RHB plus RQRPA using DD-PC1 parameterization and D1S pairing force. 
The respective $(nlj)$ quantum numbers of the SO-partner levels denote each plot 
for $\Delta E_{\rm LS}$. 
(Bottom panel) The same plot but without the D1S pairing. 
} \label{fig:2488}
\end{center} \end{figure}

\subsection{Effect of residual interactions on M1-excitation energy}
\textcolor{black}{Now we put the focus on 
how the residual interactions affect the M1-excitation energies.
Our qualitative conclusions are summarized in Table \ref{table:ry832vd}.
In the following, we present its details.
}

\textcolor{black}{In the top panel of Fig. \ref{fig:2488}, the relation between the M1-excitation and the SP SO-gap energies are presented.
For this result, the M1-peak positions ($E_{\gamma}$) are calculated using the RQRPA, whereas the SO-splitting energies ($\Delta E_{\rm LS}$) between the corresponding partner orbits are solved for the RHB quasi-particle canonical states, which are used to construct the RQRPA two-quasi-particle configuration space \cite{2003Paar, 2005Niksic}.
The analysis of the M1-excited states identifies the major SO-partner orbit(s), as denoted in the plot. 
Note that, in the RNEDF framework, the SO-gap energy results from the competition between the scalar and vector potentials \cite{1989Reinhard, 2005Vret, 2006Meng}.
}

From the top panel in Fig. \ref{fig:2488}, one can find that the actual M1 energies differ from the respective SO splittings, due to the R(Q)RPA-residual and the pairing interactions \cite{1984Borzov, 1991Migli, 1993Kam, 2009Vesely, 2010Nest, 2010Nest_2, 2019Tselyaev}. 
For example, in the $^{42}$Ca nucleus ($N=22$), the difference between its M1 energy and SO-splitting of $1f_{7/2}$-$1f_{5/2}$ levels is more than $2$ MeV. 
The similar difference commonly exists in each Ca isotope. 
The M1-SO energy difference evolves with the filling of neutron orbits along the isotope chain.
Indeed, in the top panel of Fig. \ref{fig:2488}, that difference takes the minimum value at $^{48}$Ca, where the pairing collapses in the closed shell.
Then, in open-shell nuclei, the pairing effect makes an extra contribution to expand the M1-SO difference.
There is, however, one exceptional case from this tendency: for $^{40}$Ca nucleus ($N=20$), the system turns to be M1-silent as discussed before.

For deeper knowledge, in the bottom panel of Fig. \ref{fig:2488}, we repeat the same analysis, but neglecting pairing correlations.
Namely, both of the GS and M1-excited solutions change from our default ones, because of the lack of pairing effects in the RHB and R(Q)RPA.
In this case, only the IV-PV interaction in the R(Q)RPA solution remains: see Table \ref{table:ry832vd} also.
There, the M1-excitation energies become closer to the underlying SO gaps, but still do not completely coincide.
For example, in the doubly-magic nucleus $^{48}$Ca, where the pairing correlation eventually vanishes, the difference $E_{\gamma}({\rm M1})-\Delta E_{\rm LS}$ is unchanged among the top and bottom panels in Fig. \ref{fig:2488}, and that is $E_{\gamma}({\rm M1})-\Delta E_{\rm LS}\cong 0.8$ MeV purely by the IV-PV interaction.
However, we also note that there still remain some ambiguities of the IV-PV coupling as mentioned in Sec. \ref{Sec:IVPV_Test}.
By comparing the top and bottom panels of Fig. \ref{fig:2488}, the pairing effect is shown to be important as well as that by the IV-PV interaction for the M1 mode in open-shell nuclei.

\textcolor{black}{We next mention the case, where the D1S-pairing interaction is switched on (off) in the RHB (RQRPA) calculations.
Our purpose here is to evaluate the effect of the absence of R(Q)RPA-residual interaction in the particle-particle channel.
The particle-hole IV-PV interaction is still active.
In this case, there is only a small difference found in the M1-excitation energies.
For Ca isotopes, its shift from our default case is less than $10$ keV, and thus, the result does not remarkably change from the top panel in Fig. \ref{fig:2488}.
Namely, for open-shell nuclei, the D1S-pairing interaction provides a major (minor) contribution in the RHB (RQRPA) solutions of M1.
This conclusion is equivalent to that the D1S-pairing force affects the GS properties, whereas the $1^+$-excited states are weakly sensitive to the pairing residual interaction in the RQRPA.
}
Finally, we also checked that, in the case of the unperturbed response, where both the IV-PV and pairing interactions are completely neglected in the RHB and R(Q)RPA solutions, the M1-excitation energies coincide with the SO-gap energies.

\section{Summary} \label{Sec:summary}
In this work, we have employed a relativistic multi-fermion framework to investigate the nuclear M1 excitations.
For the unnatural-parity transitions of M1 type, the RNEDF framework with the density-dependent point-coupling interaction is employed, and the corresponding RQRPA is used for the description of M1 excitations assuming the spherical symmetry.
The role of residual and pairing interactions in M1 mode has been investigated.

\textcolor{black}{Our calculations for the Ca isotope chain show that
the IV-PV coupling in terms of the R(Q)RPA residual interactions provides a finite effect on the M1 observables.
For open-shell nuclei, it is shown that the pairing correlations also play an important role to determine the M1-excitation energy and amplitude.
As the result of these interactions, although the M1 excitations are governed by transitions between SO-partner states, the excitation energy does not coincide with the respective SO splitting energy.
This conclusion is consistent to those obtained with non-relativistic approaches \cite{1984Borzov, 1991Migli, 1993Kam, 2009Vesely, 2010Nest, 2010Nest_2, 2019Tselyaev}.
We also confirmed that the present D1S-pairing correlations give a major (minor) effect on the RHB (RQRPA) solutions.
The method introduced in this work can provide a suitable way to utilize the M1 data as reference to improve the RNEDF parameters and procedures.
For this purpose, the systematic, experimental measurements of the M1 excitations are on a serious demand.
}

Before closing this paper, we note that several tasks remain for future studies.
First the quenching effect of M1 transition strength is not discussed in this study.
Indeed, in several theoretical works \cite{2010Heyde_M1_Rev, 1998VNC, 2009Vesely, 2010Nest}, 
the calculated $B({\rm M1})$ values overestimate the experimental data, when the $g$ factors of bare nucleons are used.
For adjusting the calculated $B({\rm M1})$ values, one usually needs the quenching factors that may have rather arbitrary values.
In addition, for more reliable consideration of this effect, method going beyond the standard QRPA may be required to include the advanced configurations \cite{1991Migli, 1993Kam, 1989Kamerd, 1977Dehesa, 1988Takayanagi}, similarly to the case of Gamow-Teller transitions \cite{1982Bertsch, 2006Ichimura, 2012Niu}.
Effects of the meson-exchange current and the other possible residual interactions are neither yet resolved \cite{2010Heyde_M1_Rev, 1990Richter, 2008Marcucci, 1994Moraghe}.
For neutron-rich nuclei, the deformation should be also considered.
We aim to report these improvements in future studies.

\section*{Acknowledgments}
We especially thank Tamara Nik\v{s}i\'{c} and Markus Kortelainen 
for fruitful discussions. 
This work is supported by 
the ''QuantiXLie Centre of Excellence'', a project co-financed by 
the Croatian Government and European Union through 
the European Regional Development Fund, the Competitiveness and Cohesion 
Operational Programme (KK.01.1.1.01).

\appendix
\section{\textcolor{black}{Landau-Migdal interaction} \label{App:LM_int}}
The Landau-Migdal (LM) interaction has been used in several M1 studies based on the theory of finite Fermi systems (TFFS) \cite{1984Borzov, 1991Migli, 1993Kam, 2006Speth}.
According to the M1-selection rule, the relevant terms in the LM interaction are given as
\beq
 \mathcal{F} = C_{\rm LM} \left\{ g +g'\vec{\tau}(1)\cdot \vec{\tau}(2) \right\} \bi{\sigma}(1)\cdot\bi{\sigma}(2) \delta (\bir_1 - \bir_2), \label{eq:CXTSW}
\eeq
where $\bi{\sigma}$ and $\vec{\tau}$ are the spin and isospin operators, respectively.
In Ref. \cite{1991Migli}, for example, $C_{\rm LM}=300$ MeV$\cdot$fm$^{3}$, $g=0.1$, and $g'=0.75$ are obtained from the fit to the experimental data, where $g$ is noticeably small compared with $g'$.
In the following, we show the relation between this LM interaction and the relativistic point-coupling interactions, namely its IV-PV and isoscalar-pseudovector (IS-PV) terms.

The relativistic IV-PV interaction was determined in Eq. (\ref{eq:AIVPV}) in the main text.
The corresponding interaction in the Hamiltonian is obtained as
\beqa
H_{\rm IV-PV} &=& -\int d\bir \mathcal{L}_{\rm IV-PV} \nonumber \\
&=& \hbar c\frac{\alpha_{\rm IV-PV}}{2} \int d\bir_1 \int d\bir_2  \delta(\bir_1-\bir_2) \nonumber \\
& &\psi^{\dagger}(\bir_1) \psi^{\dagger}(\bir_2)  \left[ \gamma_5(1)\cdot \gamma_5(2) +\Omega(1,2) \right] \vec{\tau}(1)\cdot \vec{\tau}(2) \psi(\bir_1) \psi(\bir_2),  \nonumber \\
\Omega(1,2)
&=& \left(  \begin{array}{cc}  \bi{\sigma}(1)\cdot\bi{\sigma}(2) &0   \\  0&\bi{\sigma}(1)\cdot\bi{\sigma}(2)  \end{array} \right).
\eeqa
In numerical calculations for its matrix elements, the first term provides the overlap integral of larger and smaller components of Dirac spinors. Thus, its contribution becomes minor than the second term. The second term then coincides with the spin-isospin term in the LM interaction in TFFS.
In the main text, we have used $\alpha_{\rm IV-PV}=0.53$ fm$^2$ as the best fit to the $^{48}$Ca and $^{208}$Pb M1 data at the RPA level. 
\textcolor{black}{In terms of the QRPA-residual interaction \cite{2003Paar}, the corresponding interaction parameter reads $\alpha_{\rm IV-PV} \hbar c=104$ MeV$\cdot$fm$^3$, which could be compared with $C_{\rm LM}g'=225$ MeV$\cdot$fm$^3$ of the LM interaction for the non-relativistic RPA \cite{1991Migli}.
Our IV-PV parameter is smaller than the LM parameter in Ref. \cite{1991Migli}, but still in a reasonable order.
Note also that, as mentioned in the section \ref{Sec:IVPV_Test}, a finite ambiguity remains in $\alpha_{\rm IV-PV}$. 
That can be attributed to 
(i) the simple-constant assumption for the parameter $\alpha_{\rm IV-PV}$, and/or 
(ii) the approximation to neglect the isoscalar-pseudovector residual interaction. 
}

The relativistic isoscalar-pseudovector (IS-PV) four-point coupling has not been employed in the present RNEDF calculations.
One reason is that the axial meson employed in the standard meson-exchange model \cite{1989Reinhard, 2005Vret, 2006Meng} is only pion, which is isovector type. 
However, that coupling is formally determined as 
\beq
\mathcal{L}_{\rm IS-PV} = -\hbar c\frac{\alpha_{\rm IS-PV}}{2} \left[ \bar{\psi}(\bir) \gamma_{5} \gamma_{\mu} \psi(\bir)  \right]  \left[ \bar{\psi}(\bir) \gamma_{5} \gamma^{\mu} \psi(\bir)  \right].
\eeq
The corresponding Hamiltonian term is represented as
\beqa
H_{\rm IS-PV} &=& \hbar c\frac{\alpha_{\rm IS-PV}}{2} \int d\bir_1 \int d\bir_2 \delta(\bir_1-\bir_2)  \nonumber \\
& & \psi^{\dagger}(\bir_1) \psi^{\dagger}(\bir_2) \left[ \gamma_5(1)\cdot \gamma_5(2) +\Omega(1,2) \right] \psi(\bir_1) \psi(\bir_2).
\eeqa
Thus, this IS-PV interaction approximately coincides with the spin energy without isospin dependence in Eq. (\ref{eq:CXTSW}).
From the analogy to the TFFS, where $g$ is shown to be minor \cite{1991Migli, 1993Kam}, we omit the IS-PV interaction in this work.
\textcolor{black}{The combined optimization of both the IV-PV and IS-PV RNEDF parameters 
is technically complicated, 
and we leave this task for future studies.
}

\section{\textcolor{black}{Non-energy-weighted sum rule of M1 strength} \label{App:KYVUT}}
We present the extended version of the M1 sum rule discussed in Ref. \cite{2019OP}.
The collective M1 excitation of the $^{A}Z$ nucleus up to the one-body-operator level is described by the following operator: 
\beq
 \oprt{Q}_{\nu} (M1) \equiv  \sum_{k \in A=N+Z} \oprt{P}^{(k)}_{\nu}(M1),
\eeq
where
\beq
 \oprt{P}^{(k)}_{\nu}(M1) = \mu_{\rm N} \sqrt{\frac{3}{4\pi}}  \left( g^{(k)}_{l} \hat{l}_{\nu}  +g^{(k)}_{s} \hat{s}_{\nu} \right).
\eeq
For simplicity, in the following, we neglect the proton's excitation, i.e. 
the system of M1-silent proton number is assumed. 
In this case, 
\beq
  \frac{1}{\mu_{\rm N}} \sqrt{\frac{4\pi}{3}} \oprt{Q}_{\nu} (M1) = g_l \hat{L}_{\nu}  +g_s \hat{S}_{\nu},
\eeq
where $\hat{L}_{\nu}=\sum_{k \in N} \hat{l}^{(k)}_{\nu}$ and $\hat{S}_{\nu}=\sum_{k \in N} \hat{s}^{(k)}_{\nu}$. 
By taking the total sum of the absolute-squared M1 amplitudes 
for $\nu =\pm1,0$ and all the excited states, 
it yields that
\beqa
  m_0 &\equiv &  \sum_{\nu} \sum_{E} \abs{\Braket{E | \left( g_l \hat{L}_{\nu}  +g_s \hat{S}_{\nu} \right) |i}}^2 \nonumber \\
  &=& \sum_{\nu} \Braket{i | \left(g_l \hat{L}_{\nu}  +g_s \hat{S}_{\nu}\right) \left(g_l \hat{L}_{\nu}  +g_s \hat{S}_{\nu}\right) |i} \nonumber \\
  &=& \Braket{i| \left(g_l \hat{\bf L}  +g_s \hat{\bf S}\right)^2 |i}. 
\eeqa
Then, by using the notation $\hat{\bf J}=\hat{\bf L}+\hat{\bf S}$, 
it is expressed as 
\beq
  m_0 = g_l(g_l-g_s) \Braket{\hat{\bf L}^2}_{[i]} +g_s(g_s-g_l) \Braket{\hat{\bf S}^2}_{[i]} +g_l g_s\Braket{\hat{\bf J}^2}_{[i]}. \label{eq:duft}
\eeq
For the GS of even-even nuclei with $J_{i}=0$, the allowed 
$(L,S)$ components must be of $L=S$, only. 
Writing this component ratio as $\abs{C_{(L,S)}}^2$, 
the summation reduces to 
\beqa
  m_0 &=& \sum_{(L,S)} \delta_{L,S} \abs{C_{(L,S)}}^2 \left\{ g_l(g_l-g_s)\cdot L(L+1) +g_s(g_s-g_l)\cdot S(S+1) \right\}  \nonumber \\
  &=& \sum_{S} \abs{C_{(L=S,S)}}^2 \left(g_l -g_s \right)^2 S(S+1). 
\eeqa
Here the total-spin number $S$ for the $N$ neutrons runs from $0$ to $N/2$, 
where $N$ was assumed as even. 
From this equation, the M1-sum value is enhanced when the high-spin components are dominant. 
On the other hand, if the pairing interaction between the valence nucleons promotes the spin-singlet configuration of valence pairs, the $S=0$ component can become dominant, and then, the M1 strength summation is expected to be small.

Note that, when both the protons and neutrons are active for M1 transitions, the M1-sum value cannot be simplified as Eq. (\ref{eq:duft}) anymore, because of the different $g$ factors. 
However, even in such a case, the qualitative conclusion does not change.
Especially, in the limit of the dominant $L=S=0$ component for protons and neutrons in even-even nuclei, the non-energy-weighted M1-sum value is zero up to the one-body-operation level.



\end{document}